\documentclass[a4paper,12pt]{article}
\usepackage[dvips]{graphicx}

\graphicspath{{figure/}}
\DeclareGraphicsExtensions{.eps}

\begin{document}
\title{ANALYTIC ESTIMATION OF THE NON-LINEAR TUNE SHIFT DUE TO THE QUADRUPOLE
MAGNET FRINGE FIELD}
\author{E.B.Levichev, P.A.Piminov,\\
Institute of Nuclear Physics SB RAS,\\Novosibirsk 630090}
\date{}
\maketitle

\begin{abstract}
Analytic expressions for the amplitude-dependent tune shift driven by the
quadrupole magnet fringe field have been obtained. The magnitude of the effect is
compared with other sources of non-linearity such as chromatic sextupoles, octupole
errors of the main quadrupole magnetic field and kinematic terms. A numerical
example was the calculation on the lattice of the \textit{VEPP-4M} $e^{+}e^{-}$ collider.
\end{abstract}

\section{Introduction}
\setcounter{equation}{0}
\setcounter{figure}{0}
\setcounter{table}{0}
The main sources of non-linear detuning for storage rings and synchrotrons
are the low-order multipole components of the magnetic field (sextupole and octupole
ones). Many authors investigated the influence of these components on beam dynamics both through
numerical simulation and analytically. The expressions estimating the non-linear betatron
tune shift for such perturbations are well-known.

There are, however, other sources of the amplitude-tune shift that are able
to influence significantly the beam behavior. For instance, the non-linearity of
the fringe field was considered in \cite{b_1} for the low-energy ring
(\textit{LER}) of the \textit{PEP-II} facility. It was
shown that the main contribution (larger than that of the regular chromatic
sextupoles) was made by the fringe field of the quadrupole magnets.
The kinematic effects and the fringe field of the bending magnets have no
significant influence.
For the \textit{B}-factory of \textit{KEK} \cite{b_2}, it is pointed out
that the main limits of the dynamical aperture are determined by the kinematic
terms of the interaction point drift space and by the fringe field of the
quadrupoles adjacent to the interaction point.

The studies are usually performed by computer simulation when the map over
the quadrupole edge is written in terms of the Lee operators \cite{b_3}
or with the help of a second-order matrix formalism \cite{b_4}.

Below is made an attempt to obtain simple analytic expressions for the
contribution of the quadrupole magnet fringe field to the non-linear tune shift.
The formulae for the kinematic effects and ordinary octupole non-linear detuning
are also presented for comparison. The octupole non-linear component is considered here as a small
error of an ideal field inside the quadrupole lens. The contribution of perturbations of
various kinds is compared on the basis of the \textit{VEPP-4M} electron-positron storage ring.

\section{Hamiltonian}
\setcounter{equation}{0}
\setcounter{figure}{0}
\setcounter{table}{0}
We will consider a quadrupole magnet with an edge field drop (see, for
instance, \cite{b_5}). We will assume that an ideal quadrupole field inside the
magnets is disturbed only by an octupole field error. Hence, we will be able
to compare its contribution to the nonlinear detuning with that induced by
the fringe field effect. The kinematic terms will also be included into the
consideration.

The transversal components of the quadrupole magnetic field are expressed as
\begin{eqnarray}
\label{m_2.1}
\nonumber
B_{z}&=&B_{1}(s)x+\frac{1}{6}B_{3}(s)(x^{3}-3xz^{2})-\frac{1}{12}B_{1}''(s)(x^{3}+3xz^{2})+O(5),\\
B_{x}&=&B_{1}(s)z+\frac{1}{6}B_{3}(s)(3x^{2}z-z^{3})-\frac{1}{12}B_{1}''(s)(z^{3}+3x^{2}z)+O(5),
\end{eqnarray}
where $B_{1}=(\partial B/\partial x)\Bigl|_{x=z=0}$ is the field gradient,
$B_{3}=(\partial^{3} B/\partial x^{3})\Bigl|_{x=z=0}$ is the octupole
component and $B''_{1}=\partial^{2} B/\partial s^{2}$
describes the contribution of the fringe field.
Since the octupole field and edge field non-linearity are presented by the same
power expansion series (but with different signs), the latter is referred to as
\textit{pseudooctupole} non-linearity sometimes.

The Hamiltonian of the transversal motion of an on-energy relativistic electron
in a magnetic field (\ref{m_2.1}) has the form of \cite{b_6}:
\begin{equation}
\label{m_2.2}
\begin{array}{lll}
H = &\frac{1}{2}(p_{x}^{2}+p_{z}^{2})+\frac{1}{2}k_{1}(s)(x^{2}-z^{2})&
\mbox{\textit{-- the undisturbed linear part,}} \\
&+\frac{1}{8}(p_{x}^{2}+p_{z}^{2})^{2}&
\mbox{\textit{-- the kinematic part,}}\\
&-\frac{1}{48}k''_{1}(s)(x^{4}+6x^{2}z^{2}-z^{4}) & \\
&-\frac{1}{2}k'_{1}(s)x^{2}zp_{z} &
\mbox{\textit{-- the influence of the fringe field,}}\\
&+\frac{1}{24}k_{3}(s)(x^{4}-6x^{2}z^{2}+z^{4}) &
\mbox{\textit{-- the octupole component,}}
\end{array}
\end{equation}
where $k_{1}(s)=B_{1}(s)/B\rho$ and $k_{3}(s)=B_{3}(s)/B\rho$.
It is considered here that $k_{1}>0$ for the horizontally-focusing quadrupole
magnet.

The main aim of this work is to investigate the non-linearity of the fringe field.
However, for the purpose of comparison and estimation, expressions for the kinematic
terms and for the regular octupole component of the magnetic field will be also
obtained.

In order to find the amplitude-tune dependence in the first order of
approximation, let us write down expressions (\ref{m_2.2}) in the "
"action-angle" variables \cite{b_7} with the help of the generating function
\begin{equation}
\label{m_2.3}
F(x,z,\phi_{x},\phi_{x},s)=
-\frac{z^{2}}{2\beta_{z}(s)}\Biggl[\tan\phi_{z}-\frac{\beta^{\prime}_{z}(s)}{2}\Biggr]
-\frac{x^{2}}{2\beta_{x}(s)}\Biggl[\tan\phi_{x}-\frac{\beta^{\prime}_{x}(s)}{2}\Biggr],
\end{equation}
which assigns the following relations for the old and  new variables:
\begin{equation}
\label{m_2.4}
\begin{array}{lll}
x=\sqrt{2J_{x}\beta_{x}(s)}\cos\phi_{x},&
p_{x}=-\sqrt{\frac{2J_{x}}{\beta_{x}(s)}}\Bigl(\sin\phi_{x}+\alpha_{x}(s)\cos\phi_{x}\Bigr),&\\
z=\sqrt{2J_{z}\beta_{z}(s)}\cos\phi_{z},&
p_{z}=-\sqrt{\frac{2J_{z}}{\beta_{z}(s)}}\Bigl(\sin\phi_{z}+\alpha_{z}(s)\cos\phi_{z}\Bigr),&\\
\end{array}
\end{equation}
where $\alpha$ and $\beta$ are the Twiss parameters.

Averaging the new Hamiltonian over all phase variables,
$\langle H(s)\rangle_{\phi_{z},\phi_{x}}$, and computing the oscillation
tune of the system according to
\begin{equation}
\label{m_2.5}
\nu_{x,z}=\frac{1}{2\pi}\oint\frac{\partial}{\partial J_{x,z}}\langle H(s)\rangle_{\phi_{z},\phi_{x}}ds,
\end{equation}
we obtain the following first-order expressions for the amplitude-tune
dependence:
\begin{eqnarray}
\label{m_2.6}
\nonumber
\Delta\nu_{x}&=&C_{xx}J_{x}+C_{xz}J_{z},\\
\Delta\nu_{z}&=&C_{xz}J_{x}+C_{zz}J_{z},
\end{eqnarray}
where the coefficients include the contribution of the kinematic effects, fringe field
and regular octupole perturbation: $C=C^{k}+C^{e}+C^{o}$. Coefficients of
each type have the following form:
\begin{enumerate}
\item[a)] Kinematic coefficients.
\begin{displaymath}
C_{xx}^{k}=\frac{3}{16\pi}\oint\gamma_{x}^{2}(s)ds,~
C_{zz}^{k}=\frac{3}{16\pi}\oint\gamma_{z}^{2}(s)ds,
\end{displaymath}
\begin{equation}
\label{m_2.7}
C_{xz}^{k}=\frac{1}{8\pi}\oint\gamma_{x}(s)\gamma_{z}(s)ds,
\end{equation}
where $\gamma=\bigl(1+\alpha^{2}\bigr)/\beta$.
\item[b)] Fringe field coefficients.
\begin{equation}
\label{m_2.8}
C_{xx}^{e}=-\frac{1}{32\pi}\oint k_{1}^{\prime\prime}(s)\beta_{x}^{2}(s)ds,~
C_{zz}^{e}=+\frac{1}{32\pi}\oint k_{1}^{\prime\prime}(s)\beta_{z}^{2}(s)ds,
\end{equation}
\begin{displaymath}
C_{xz}^{e}=-\frac{1}{16\pi}\oint\beta _{x}(s)
\Bigl( k_{1}^{\prime\prime}(s)\beta_{z}(s)-4k_{1}^{\prime}(s)\alpha_{z}(s)\Bigr)ds.
\end{displaymath}
\item[c)] Regular octupole component coefficients.
\begin{equation}
\label{m_2.9}
C_{xx}^{o}=\frac{1}{16\pi}\oint k_{3}(s)\beta_{x}^{2}(s)ds,
C_{zz}^{o}=\frac{1}{16\pi}\oint k_{3}(s)\beta_{x}^{2}(s)ds,
\end{equation}
\begin{displaymath}
C_{xz}^{o}=\frac{1}{8\pi}\oint k_{3}(s)\beta_{x}(s)\beta_{z}(s)ds.
\end{displaymath}
\end{enumerate}

For the sake of clearness and comparison, let us simplify general expressions
(\ref{m_2.7})--(\ref{m_2.9}) basing on the evaluation of each type of
perturbation in a cyclic accelerator.

\section{Kinematic effects}
\setcounter{equation}{0}
\setcounter{figure}{0}
\setcounter{table}{0}
The main contribution to integral expressions (\ref{m_2.7}) is made by the
drift space with an extremely small value of the betatron function. The
interaction region is best suited to consideration from such point
of view (see, for instance, \cite{b_2}). For a straight section with the
length $L$, where behavior of the beta-functions is mirror-symmetric
relative to the center, $\gamma=1/\beta_{0}=const$ ($\beta_{0}$ is the value
at the center of the section), expressions (\ref{m_2.7}) take a simple form,
convenient for estimation:
\begin{equation}
\label{m_3.1}
C_{xx}^{k}=\frac{3}{16\pi}\frac{L}{\beta_{0x}^{2}},~
C_{xz}^{k}=\frac{1}{8\pi}\frac{L}{\beta_{0x}\beta_{0z}},~
C_{zz}^{k}=\frac{3}{16\pi}\frac{L}{\beta_{0z}^{2}}.
\end{equation}

\section{Octupole errors}
\setcounter{equation}{0}
\setcounter{figure}{0}
\setcounter{table}{0}
If the magnet lattice of the accelerator is known and, thus, the behavior of the
beta-functions along the beam trajectory is calculated, one can easily compute
coefficients (\ref{m_2.9}) numerically. However, we obtain the
following simple estimation to be able to compare the
influence of the octupole error of the central part of a quadrupole lens $L$
long with the influence of the fringe fields of this lens.

Let us assume, for the coefficient $C^{o}_{xx}$, that $\beta_{x}$ is approximately
constant inside the lens ($k_{1}\ll 1$) and the octupole component $k_{3}$
is constant over the length, then
\begin{equation}
\label{m_4.1}
C_{xx}^{o}\approx\frac{1}{16\pi}k_{3}\beta_{0x}^{2}L.
\end{equation}

The quality $q$ of a magnetic field is usually defined as the relative difference
between the real and ideal values of the field at the radius $a$ (we will regard
this radius as the inscribed radius of the aperture of the quadrupole lens).
Then, from (\ref{m_2.1})  for the field $B_{z}$ in the median plane
($z=0$) it follows that
\begin{equation}
\label{m_4.2}
q=\frac{k_{3}}{6k_{1}}\cdot a^{2}.
\end{equation}
Substituting (\ref{m_4.2}) into (\ref{m_4.1}), we obtain the estimation
 we will need later:
\begin{equation}
\label{m_4.3}
C_{xx}^{o}\approx\frac{3}{8\pi a^{2}}k_{1}q\beta_{0x}^{2}L.
\end{equation}

\section{Fringe fields}
\setcounter{equation}{0}
\setcounter{figure}{0}
\setcounter{table}{0}
The non-linear betatron tune shift caused by the fringe fields of lenses is
the principal subject of this work. In order to simplify expression
(\ref{m_2.8}), we need to introduce a model describing the behavior of the
fringe field of the quadrupole lens.

We have considered the following variants of model distribution of the
quadrupole lens fringe field: the one based on M.Bassettis's formulae
\cite{b_8, b_9}, the model of \textit{"four lines"}" by \textit{G.Lee-Whiting}
\cite{b_10} and the description of the magnetic field drop with two matched
parabolas \cite{b_11}. The first two models describe rather precisely the
fringe field and its derivatives with respect to the longitudinal coordinate.

\begin{figure}
\centering
\includegraphics[width=0.9\textwidth]{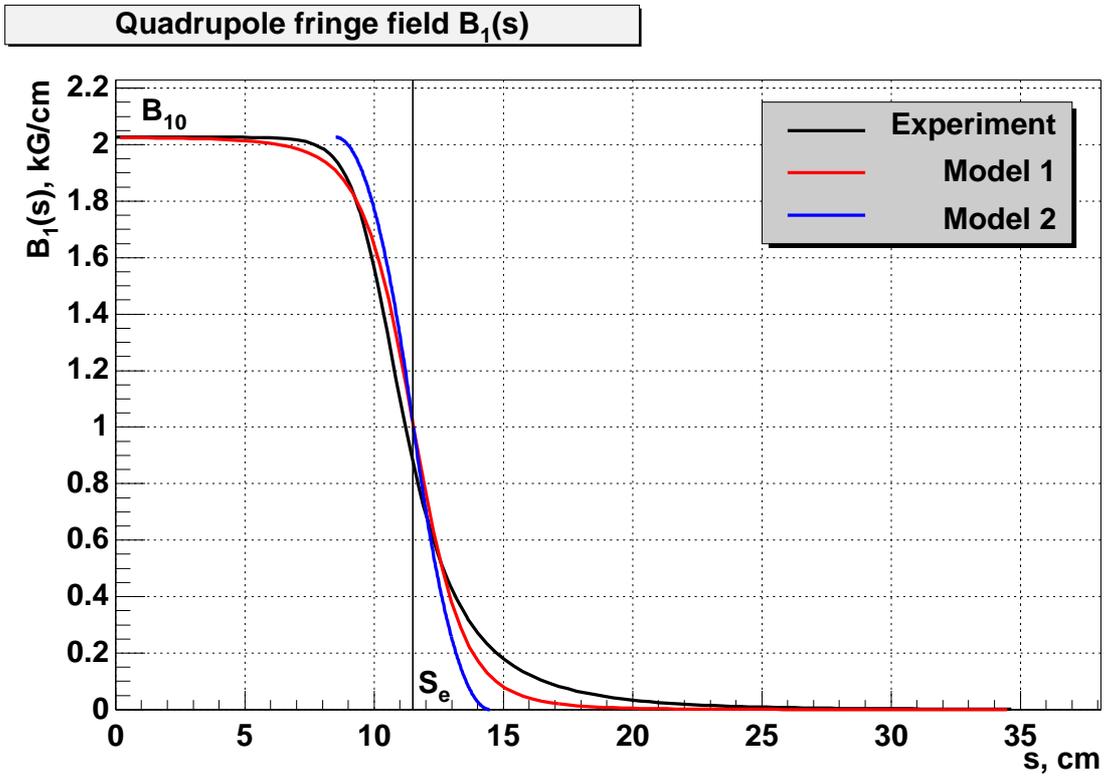}
\caption{Fringe gradient drop: measured and computed,
according to (\ref{m_5.1}) (model 1) and according to (\ref{m_5.2}) (model 2).}
\label{f_5.1a}
\end{figure}

\begin{figure}
\centering
\includegraphics[width=0.9\textwidth]{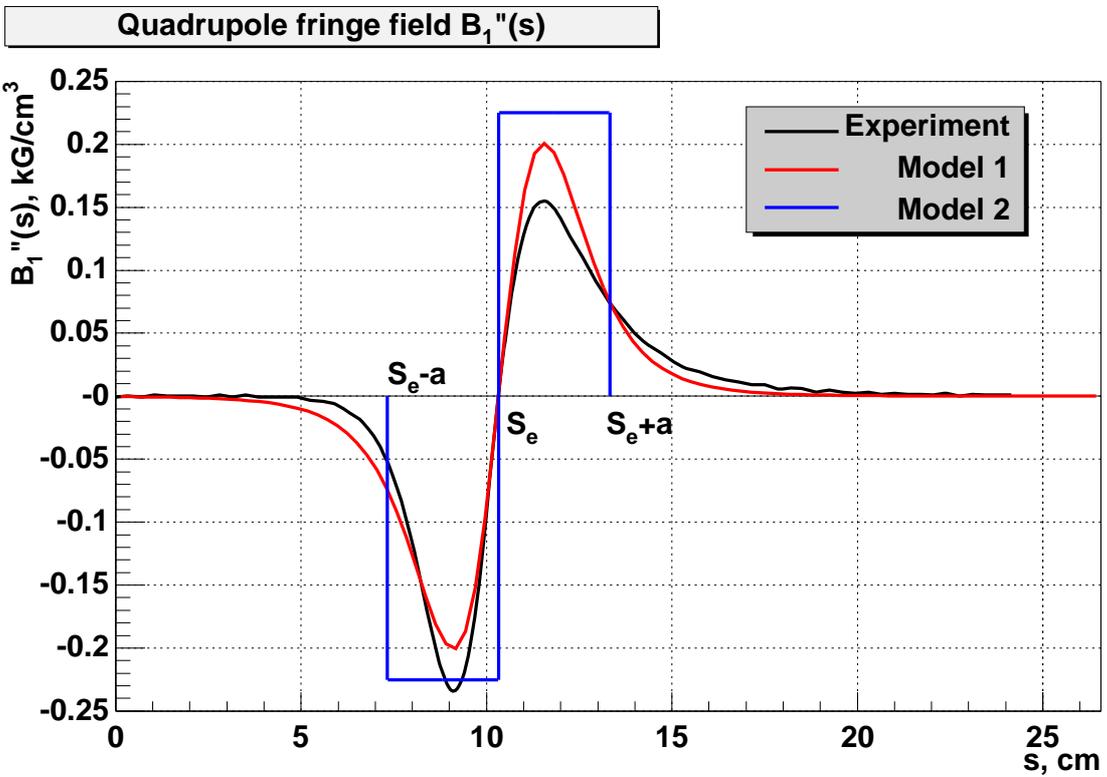}
\caption{Second derivative of the gradient with respect to the
longitudinal coordinate. Notations are in Fig.\ref{f_5.1a}.}
\label{f_5.1b}
\end{figure}

For instance, for the \textit{"four-line"} model, the distribution of the gradient
and its first two derivatives (required for computation of
(\ref{m_2.8})) at the edge of the lens has the following form:
\begin{eqnarray}
\label{m_5.1}
\nonumber
B_{1}(s)&=&\frac{1}{2}B_{10}\Biggl(1-\frac{3}{2}\frac{s}{d}+\frac{1}{2}\frac{s^{3}}{d^{3}}\Biggr),\\
\nonumber
B_{1}^{\prime}(s)&=&-\frac{3}{4}B_{10}\frac{1}{d}\Biggl(1-\frac{s^{2}}{d^{2}}\Biggr)^{2},\\
B_{1}^{\prime\prime}(s)&=&\frac{15}{4}B_{10}\frac{s}{d^{3}}\Biggl(1-\frac{s^{2}}{d^{2}}\Biggr)^{2},\\
\nonumber
&&d=\sqrt{a^2+s^2},
\end{eqnarray}
where $a$ is the inscribed radius of the aperture of the lens and $B_{10}$ is
the gradient on the lens axis in the central section. The approximation of the
fringe field with the help of (\ref{m_5.1}) is shown in Fig.\ref{f_5.1a}, \ref{f_5.1b}.
Unfortunately, it is rather complicated to use (\ref{m_5.1}) since the field
distribution is assumed infinite on each side of the axis (the edge
corresponds to $s=0$). So, a correct integration of (\ref{m_2.8}) with due
regard to the real behavior of the betatron function becomes difficult.

At the same time, a simple model of two matched parabolas \cite{b_1}
\begin{equation}
\label{m_5.2}
B_{1}(s)=\Biggl\{
\begin{array}{ll}
\frac{B_{10}}{2a^{2}}\bigl(a+s\bigr)^{2},&a\leq s\leq 0\\
B_{10}-\frac{B_{10}}{2a^{2}}\bigl(a-s\bigr)^{2},&a\geq s\geq 0\\
\end{array}
\end{equation}
yields a rather good approximation of the resulting expressions without
cumbersome evaluation of numeric coefficients for the model
(\ref{m_5.1}). The difference between the value of coefficients for models
(\ref{m_5.1}) and (\ref{m_5.2}) will be shown below.

Fig.\ref{f_5.1a}, \ref{f_5.1b} present the measured gradient drop of the quadrupole
lens \cite{b_12} and the model expression according to (\ref{m_5.1}) and
(\ref{m_5.2}). The main notations used below are also given. The inscribed radius
of the aperture of the lens $a=2{.}8$ cm; the maximal gradient at the center
of the lens $B_{10}\cong 20$ T/m. It is worth noting that because of the
quadrupole symmetry the end-pole chamfer, usually used for adjustment of the
efficient length of the lens, has practically no influence on the fringe
field pseudooctupole component defined by $B^{\prime}_{1}(s).$

The fringe effects are concentrated in the vicinity of the longitudinal edge
of the lens with a typical region length close to the lens inscribed radius:
$\sim\pm a$. We assume that the betatron functions vary a little over this
length, so they can be presented in the first-order expansion. For instance,
for the horizontal beta-function:
\begin{equation}
\label{m_5.3}
\beta_{x}(s)\approx\beta_{x_{e}}+\beta_{x_{e}}^{\prime}\Bigl(s-s_{e}\Bigr),
\end{equation}
where $\beta_{x_{e}}$ and $\beta_{x_{e}}^{\prime}$ are the magnitudes of the
beta-function and its derivative at the edge and $e=1,2$ for the first and
second edges.

Let us substitute (\ref{m_5.3}) and corresponding derivatives of the fringe field
(\ref{m_5.2}) into (\ref{m_2.8}) and perform the integration. Then we can obtain
the following expressions describing the contribution of the fringe field of
the quadrupole lens to the non-linear shift of the betatron tune:
\begin{eqnarray}
\label{m_5.4}
\nonumber
C^{e}_{xx}&=&\frac{1}{16\pi}k_{10}\bigl(\beta_{x_{1}}\beta_{x_{1}}^{\prime}-\beta_{x_{2}}\beta_{x_{2}}^{\prime}\bigr),\\
\nonumber
C^{e}_{xz}&=&\frac{1}{16\pi}k_{10}\bigl(\beta_{z_{1}}\beta_{x_{1}}^{\prime}-\beta_{x_{1}}\beta_{z_{1}}^{\prime}-\beta_{z_{2}}\beta_{x_{2}}^{\prime}+\beta_{x_{2}}\beta_{z_{2}}^{\prime}\bigr),\\
C^{e}_{zz}&=&\frac{1}{16\pi}k_{10}\bigl(\beta_{z_{1}}\beta_{z_{1}}^{\prime}-\beta_{z_{2}}\beta_{z_{2}}^{\prime}\bigr),
\end{eqnarray}
where $k_{10}$ is the focusing coefficient of the lens at the center. It is
taken with its proper signs for the focusing and defocusing lenses. It is
worth noting that the expressions obtained include only the parameters of the linear optics
 and do not include the lens aperture $a$.  The complete value of the
non-linear dependence of the tune on the amplitude is obtained by summing (\ref{m_5.4})
over all the quadrupole lenses of the accelerator.

Usage of more accurate distribution of the fringe field (\ref{m_5.1}) and
integration of (\ref{m_2.8}) close to the edge of the lens in the interval
$\pm na$ gives the following correction in (\ref{m_5.4}): $k_{10}$ should be
replaced with
\begin{displaymath}
\frac{k_{10}}{2}\frac{n^{3}}{\bigl(n^{2}+1\bigr)^{3/2}}\Biggl(5-3\frac{n^{2}}{n^{2}+1}\Biggr),
\end{displaymath}
i.e. with $n=2$ the difference is 7\% and with $n=3$ it is as small as 2\%.

Formulae (\ref{m_5.4}) can be easily used numerically if the distribution of the
betatron functions along the ring is available. However, to determine the
general features of the influence of the quadrupole fringe field upon the
non-linear shift of the betatron tune, it would be useful to simplify (\ref{m_5.4}) still
further assuming that $\sqrt{k_{10}}L<1$ for the quadrupole lens under consideration. Then
(\ref{m_5.4}) can be expanded in series to the second order inclusive.
\begin{eqnarray}
\label{m_5.5}
\nonumber
C^{e}_{xx}&\approx&\frac{1}{8\pi}k_{10}^{2}\beta_{x0}^{2}L\bigl(1-\frac{1}{4}k_{10}L^{2}\bigr),\\
\nonumber
C^{e}_{xz}&\approx&\frac{1}{4\pi}k_{10}^{2}\beta_{x0}\beta _{z0}L,\\
C^{e}_{zz}&\approx&\frac{1}{8\pi}k_{10}^{2}\beta_{z0}^{2}L\bigl(1+\frac{1}{4}k_{10}L^{2}\bigr),
\end{eqnarray}
where $\beta_{0}$ is the extreme (maximum or minimum) value of the
corresponding betatron function achieved at the center of the lens. The last
assumption is not essential and is made only because of the especially simple
form of formulae (\ref{m_5.5}) in that case. In the case of arbitrary behavior of
the betatron function (still having extremes inside the lens), $L^{2}/4$
in the brackets should be replaced with $l_{1}^{2}-l_{1}l_{2}+l_{2}^{2}$,
where $l_{1,2}$ is the distance from the lens edges to the extreme of the
betatron function.

One can see from (\ref{m_5.5}) that the frequency-amplitude dependence is
always positive for the fringe field of the lens (at least, for the first-order term of the expansion),
be this lens a focusing or a defocusing
one. The last fact is illustrated well by Table \ref{t_6.1} obtained
numerically for \textit{LER} \cite{b_1}.

Let us compare the influence of the octupole error of the field inside the lens with
that of the pseudo-octupole fringe component.  From (\ref{m_4.3})
and (\ref{m_5.5}) one can obtain:
\begin{equation}
\label{m_5.6}
\frac{C^{e}_{xx}}{C^{o}_{xx}}\approx\frac{k_{10}a^{2}}{3q}.
\end{equation}

Substituting reasonable parameters of the quadrupole lens in (\ref{m_5.5}),
one can see that the influence of the fringe field is comparable with and can even
exceed the field errors in the inner area.

\section{Comparison of different contributions to the non-linear
tune shift}
\setcounter{equation}{0}
\setcounter{figure}{0}
\setcounter{table}{0}
\begin{figure}
\centering
\includegraphics[width=\textwidth]{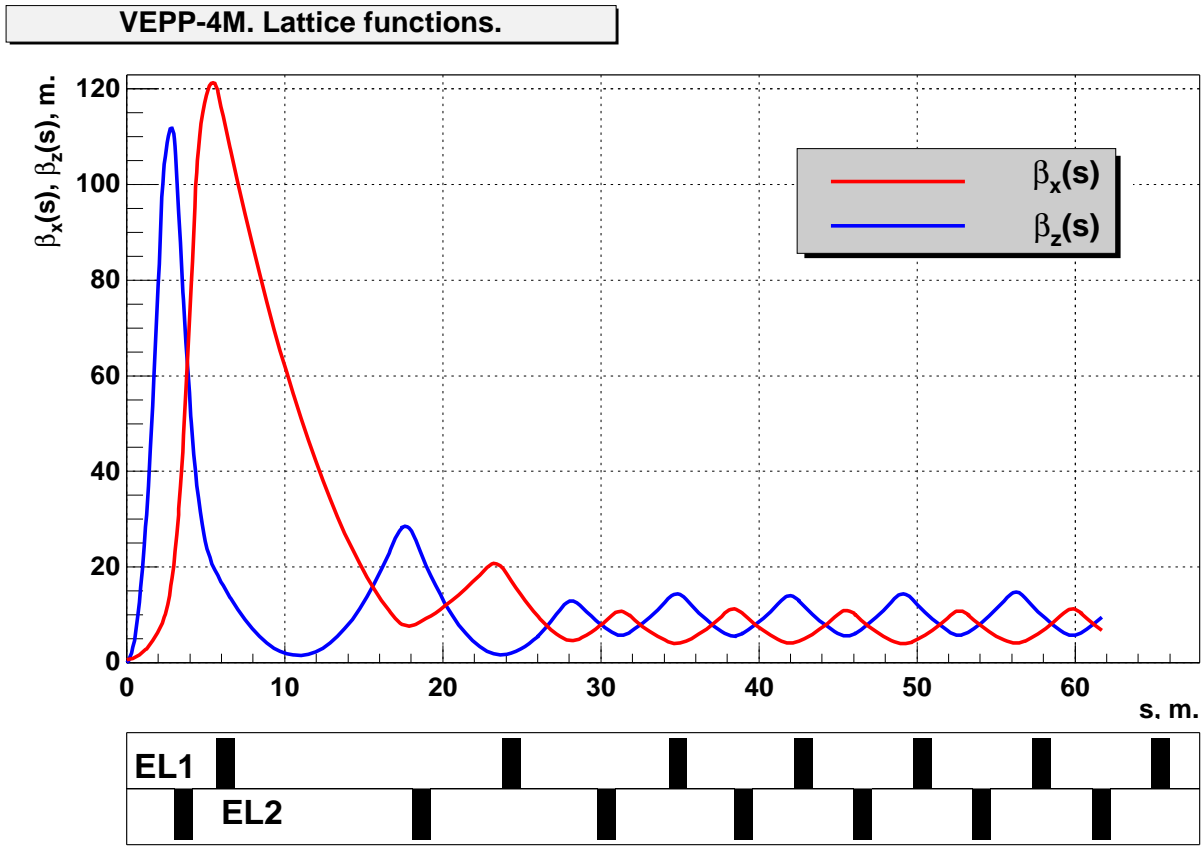}
\caption{Behavior of the lattice functions in the region of the \textit{VEPP-4M} final
focus.}
\label{f_6.1}
\end{figure}

The magnet lattice of the electron-positron storage ring of \textit{VEPP-4M}
\cite{b_13} was used as a model; the following comparisons were made for this
machine:
\begin{itemize}
\item The magnitudes of the non-linear shift of the betatron tune for perturbation of
different kinds, including the fringe field, octupole error of field of the
lenses, kinematic effects and sextupole magnets, compensating the natural
chromaticity;
\item The tune dependence on the amplitude for the fringe field of the quadrupole
lenses (obtained by the numerical modeling and in the analytic form, with the
use of (\ref{m_5.4}) and (\ref{m_5.5})).
\end{itemize}

The octupole component of the field of the quadrupole lenses was taken from
the results of the magnetic measurements \cite{b_12}. Its most significant
value was in the defocusing lens of the final focus (\textit{EL2} in
Fig.\ref{f_6.1}).

From the magnetic mapping,  $B_{3}\cong 0{.}2$ G/cm$^{3}$ for this lens with an
inscribed aperture diameter of 170 mm at an energy of $1{.}85$ GeV, for which
the calculations were performed. For the rest lenses of the ring this value is
from 7 to 10 times less.

Table \ref{t_6.1} presents the comparative data for different sources of
non-linearity for \textit{VEPP-4M}
\begin{figure}
\centering
\includegraphics[width=\textwidth]{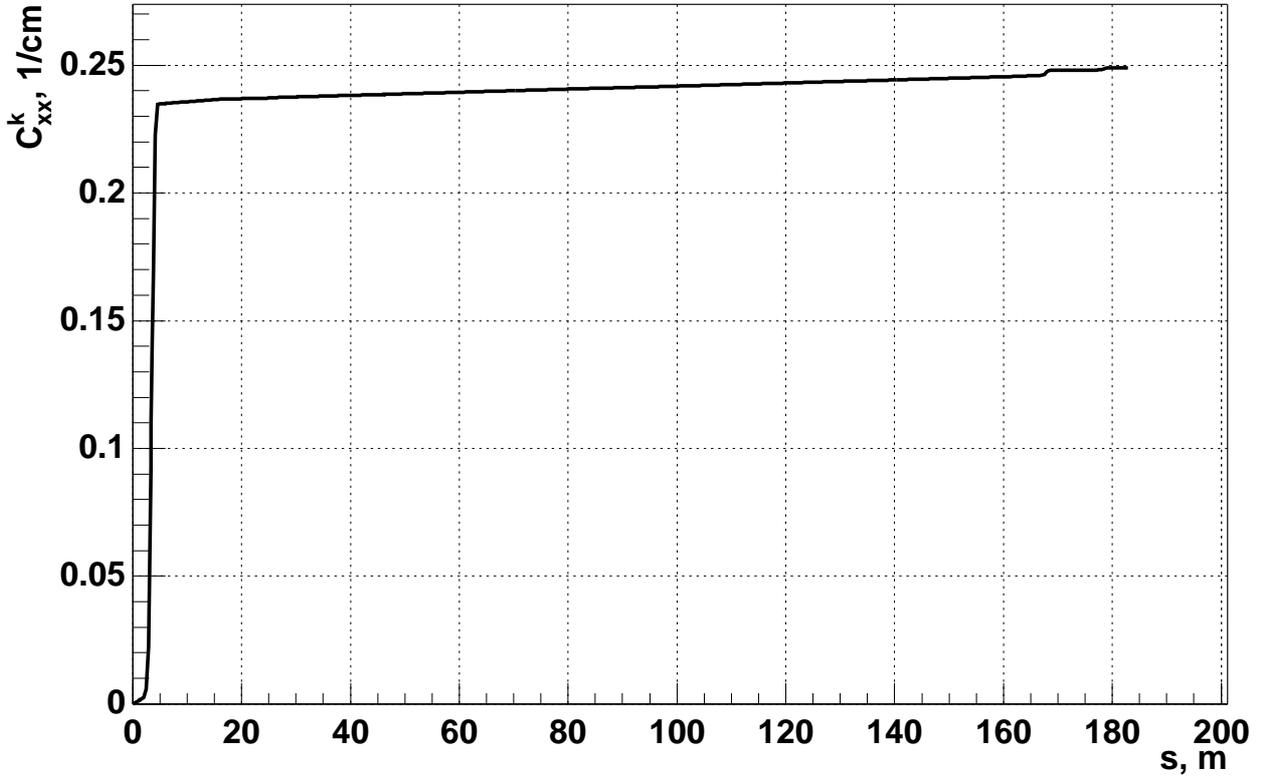}
\caption{$C_{xx}$ due to the kinematic effects for $1/2$ of the \textit{VEPP-4M} ring.
The interaction point corresponds to $s=0$.}
\label{f_6.2}
\end{figure}
\begin{table}
\centering
\caption{Values of coefficients (\ref{m_2.6}) for \textit{VEPP-4M}.}
\label{t_6.1}
\begin{tabular}{|l|c|c|c|}
\hline 				& $C_{zz}$, (m$^{-1}$)	& $C_{xz}$, (m$^{-1}$)	& $C_{zz}$, (m$^{-1}$)	\\
\hline Kinematic effects	& 50			& 7 			& 96			\\
\hline Fringe field		& 170			& 300			& 390			\\
\hline Sextupole lenses		& 88			& -1680			& -1660			\\
\hline Octupole errors		& -510			& -340			& -280			\\
\hline
\end{tabular}
\end{table}

It is seen that for the case of \textit{VEPP-4M} the influence of the fringe field is
relatively small. However, the corresponding values obtained for \textit{LER}
\cite{b_1} by the numerical simulation exceed  the \textit{VEPP-4M}
values several times and their contribution to the non-linear shift is the most important
among all kinds of perturbation. For instance, $C_{xx}$ from the sextupoles
equals –800 m$^{-1}$ while that from the fringe field effects is 1628
m$^{-1}$.

\begin{figure}
\centering
\includegraphics[width=\textwidth]{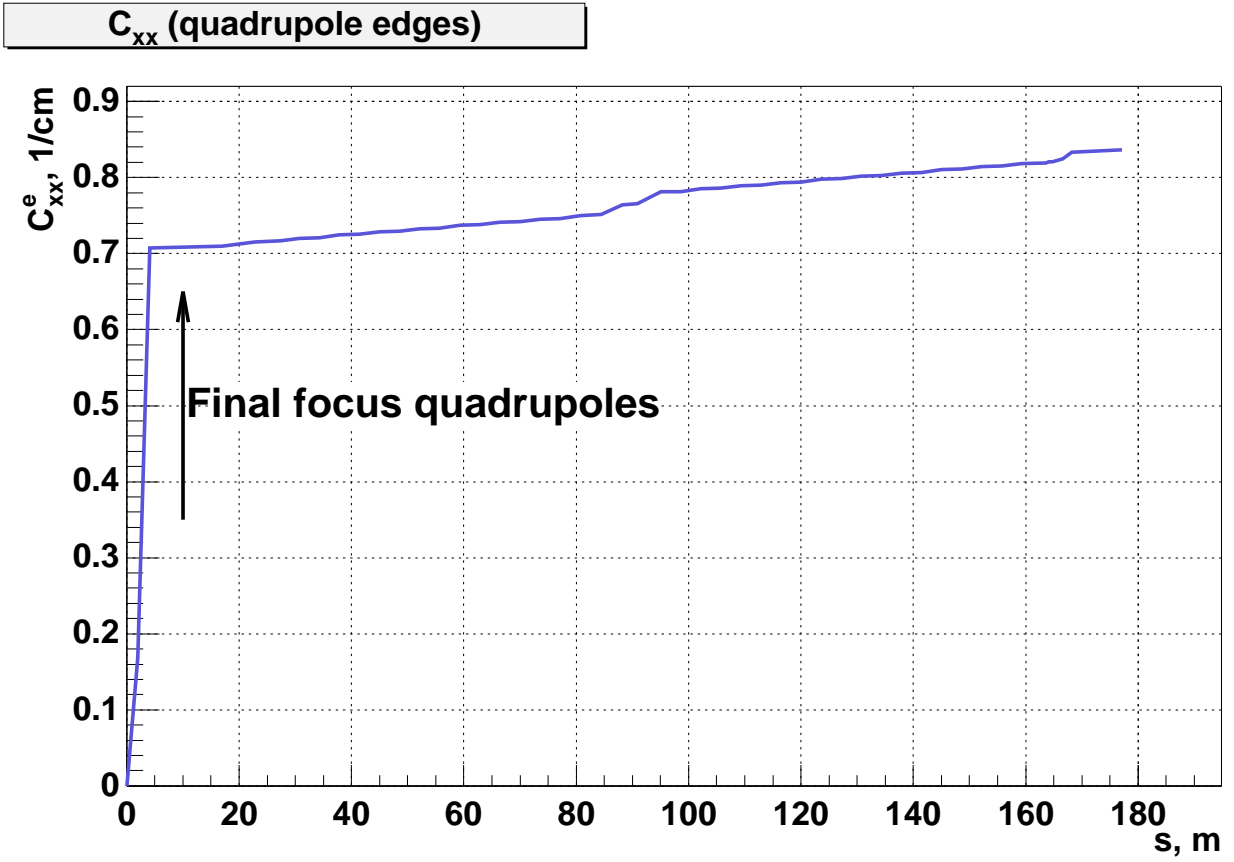}
\caption{$C_{xx}$ due to the fringe effects for $1/2$ of the \textit{VEPP-4M} ring.}
\label{f_6.3}
\end{figure}

In case of kinematic effects, as expected, the main contribution is made by
the interaction drift space (Fig.\ref{f_6.2}). The quadrupole lenses
\textit{EL1} and \textit{EL2} of the final focus respond mainly to the fringe
field non-linearity because the beta-functions here are from
110 to 120 m (Fig.\ref{f_6.3}).
\begin{table}
\centering
\caption{Numerical simulation and analytical evaluation of the fringe effects}
\label{t_6.2}
\begin{tabular}{|l|c|c|c|}
\hline					& $C_{zz}$, (m$^{-1}$)	& $C_{xz}$, (m$^{-1}$)	& $C_{zz}$, (m$^{-1}$)	\\
\hline Simulation			& 182			& 308			& 403			\\
\hline Evaluation by (\ref{m_5.4})	& 170			& 300			& 390			\\
\hline Evaluation by (\ref{m_5.5})	& 170			& 260			& 350			\\
\hline
\end{tabular}
\end{table}

In order to compare the evaluation of the fringe field with the numerical
simulation, the latter was carried out for the case of a "piecewise"
approximation of the quadrupole lens field. In so doing, according to
\cite{b_3, b_14}, both the particle coordinate and the momentum change stepwise at
the edge of the lens:
\begin{eqnarray}
\label{m_6.1}
\nonumber
\Delta x&=&\frac{1}{12}k_{10}\bigl(x^{3}+3xz^{2}\bigr),\\
\Delta p_{x}&=&\frac{1}{4}k_{10}\Bigl[2xzp_{z}-p_{x}\bigl(x^{2}+z^{2}\bigr)\Bigl].
\end{eqnarray}
For the vertical plane the substitutions  $x\leftrightarrow z$ and
$k_{10}\leftrightarrow -k_{10}$ are made. The results of the simulation and
analytical evaluation are compared in Table \ref{t_6.2}.

It is seen that the analytic evaluation differs from the simulation by several
percent. This can be explained by the accuracy of extraction of the tune shift
at the Fourier discrete analysis of 1024 revolutions of the tracking (just
a $\sim 10\%$ level of the absolute value of the non-linear shift).

So, the evaluation of the dependence of the betatron tune on the amplitude for the
fringe fields of the quadrupole lenses agrees rather satisfactorily with the
results of the numerical simulation. Thus, one can obtain the result quickly
from the linear magnetic structure of the cyclic accelerator, especially as
few codes of the numerical tracking allow taking into account the
influence of fringe non-linearity for the simulation.

\section{Acknowledgements}
The authors are grateful to S.F.Mikhailov for granting the stuff of the
magnetic measurements and to I.Ya.Protopopov for assistance in the mathematical
simulation of the fringe effects.

\end{document}